\documentclass[twocolumn,prl,aps,superscriptaddress]{revtex4-1}
\usepackage[latin9]{inputenc}
\usepackage{mathtools}
\usepackage{amsbsy}
\usepackage{amstext}
\usepackage{appendix}
\usepackage{braket}
\usepackage{multirow}
\usepackage{array}
\usepackage[unicode=true,pdfusetitle,
 bookmarks=false,
 breaklinks=false,pdfborder={0 0 1},backref=false,colorlinks=false]
 {hyperref}
\usepackage[table]{xcolor}
\usepackage{changepage}
\usepackage{dsfont}
\usepackage{colortbl}

\definecolor{lightgray}{gray}{0.9}

\usepackage[english]{babel}

\hypersetup{
 bookmarksnumbered=false,bookmarksopen=false}
\makeatletter

\usepackage{amsmath}
\usepackage{graphicx}
\usepackage{amssymb}
\usepackage{txfonts,color}
\@ifundefined{textcolor}{}

{%
 \definecolor{BLACK}{gray}{0}
 \definecolor{WHITE}{gray}{1}
 \definecolor{RED}{rgb}{1,0,0}
 \definecolor{GREEN}{rgb}{0,1,0}
 \definecolor{BLUE}{rgb}{0,0,1}
 \definecolor{CYAN}{cmyk}{1,0,0,0}
 \definecolor{MAGENTA}{cmyk}{0,1,0,0}
 \definecolor{YELLOW}{cmyk}{0,0,1,0}
 \definecolor{ORANGE}{rgb}{1,0.5,0}
}

\makeatother

\begin{document}
\author{P. Alsina-Bol\'{i}var}
\affiliation{Department of Physical Chemistry, University of the Basque Country UPV/EHU, Apartado 644, 48080 Bilbao, Spain}
\affiliation{EHU Quantum Center, University of the Basque Country UPV/EHU, Leioa, Spain}
\author{I. Iriarte-Zendoia}
\affiliation{Department of Physical Chemistry, University of the Basque Country UPV/EHU, Apartado 644, 48080 Bilbao, Spain}
\affiliation{EHU Quantum Center, University of the Basque Country UPV/EHU, Leioa, Spain}
\author{D. B. Bucher}
\affiliation{Technical University of Munich, TUM School of Natural Sciences, Department of Chemistry, Lichtenbergstra{\ss}e 4, Garching bei M{\"u}nchen, 85748, Germany}
\affiliation{Munich Center for Quantum Science and Technology (MCQST), Schellingstr. 4, M{\"u}nchen, 80799, Germany}
\author{J. Casanova}
\affiliation{Department of Physical Chemistry, University of the Basque Country UPV/EHU, Apartado 644, 48080 Bilbao, Spain}
\affiliation{EHU Quantum Center, University of the Basque Country UPV/EHU, Leioa, Spain}

\title{Nanoscale Sensing of Solid-State Samples with High Frequency Resolution}

\begin{abstract}
To meet the growing demand for nanoscale surface analysis, nitrogen-vacancy (NV) centers offer a high-sensitivity alternative by leveraging their ability to operate in immediate proximity to the sample. In this work, we propose a quantum control protocol designed to overcome the inherent challenges of solid-state environments, specifically by mitigating anisotropy and strong dipole-dipole interactions to enable the detection of isotropic chemical shifts at the nanoscale. To achieve this, our scheme synchronizes a slowly rotating magnetic field with tailored RF decoupling and MW control of the NV sensors. We provide an analytical mapping that explicitly links the measured spectrum to the control sequence features and the underlying system parameters, enabling a straightforward characterization of the sample.
\end{abstract}

\maketitle
Over the past decade, solid-state quantum sensors have emerged as a transformative platform to conventional probing techniques, particularly nuclear magnetic resonance (NMR), as their close proximity to the sample led to excellent volume sensitivities~\cite{allert2022advances,rizzato2023quantum,du2024single}.  This capability aligns with one of the central goals of analytical science: determining molecular structure and composition. This is relevant in chemistry, biology, and materials science, and is typically addressed by methods such as X-ray crystallography, mass spectrometry, atomic force microscopy, and NMR~\cite{cerofolini2019integrative,ziegler2021advances,sikic2010systematic,jiang2025structural}. 

In particular, nitrogen vacancy (NV) centers in diamond~\cite{doherty2013the} are approaching competitive performance for liquid-state analysis at room temperature \cite{glenn2018high,bucher2020hyperpolarization,arunkumar2021micronscale,staudacher2013nuclear,mamin2013nanoscale,schmitt2017submillihertz,boss2017quantum,grafenstein2025coherent,munuera2023high,munuera2022detection,biteri2023amplified}. At the microscale, NV ensemble detection  report high spectral resolution, beyond hertz, and sensitivities around $\mathrm{pT\,Hz^{-1/2}}$ by exploiting quantum heterodyne detection schemes~\cite{glenn2018high,schmitt2017submillihertz,boss2017quantum,munuera2023high,alsina2024jcoupling,whaites2025high}. For nanoscale samples, shallowly implanted NV centers provide nanometre-scale spatial resolution and sensitivity volumes on the order of \( (5\ \mathrm{nm})^3\)~\cite{mamin2013nanoscale,staudacher2013nuclear,lovchinsky2017magnetic,aslam2017nanoscale,schwartz2019blueprint,spohn2025quantum}. However, in this scenario molecular diffusion leads to poor spectral resolution, which typically range from kHz to MHz~\cite{Staudenmaier2022power,liu2022using}.

The interrogation of solid-state samples is of great interest across many fields, including biomaterials~\cite{tycko2011solid}, pharmaceutics and drug delivery~\cite{pisklak201613c,marchetti2021solid}, surface analysis~\cite{tycko2011solid,liu2022surface}, and renewable energy and energy-storage research~\cite{pecher2017materials,addison2025solid}. In this scenario, the absence of molecular diffusion is advantageous in terms of  frequency resolution, however dipolar couplings and  chemical shift anisotropy (CSA) are no longer averaged out which results in challenging spectra. Conventional solid-state NMR techniques mitigate such interactions through magic-angle spinning (MAS)~\cite{andrew1958nuclear,duer2008solid,nishiyama2022ultrafast}, where the sample is rapidly rotated. MAS is often paired with schemes like Lee-Goldburg (LG) irradiation~\cite{lee1965nuclear} to suppress homonuclear dipolar couplings at moderate rotation speeds. This simplifies the experimental setup and minimizes MAS-related frequency artifacts.

Techniques that leverage quantum sensing for solid-state detection have been proposed with a focus on microscale ordered solids~\cite{munuera2025pulse}. However, there is growing interest in characterizing critical systems --such as material surfaces and thin films-- at the nanoscale. Achieving this remains challenging due to the interplay between chemical shift anisotropy and dipolar interactions.

\begin{figure*}[]
\includegraphics[width=1 \linewidth]{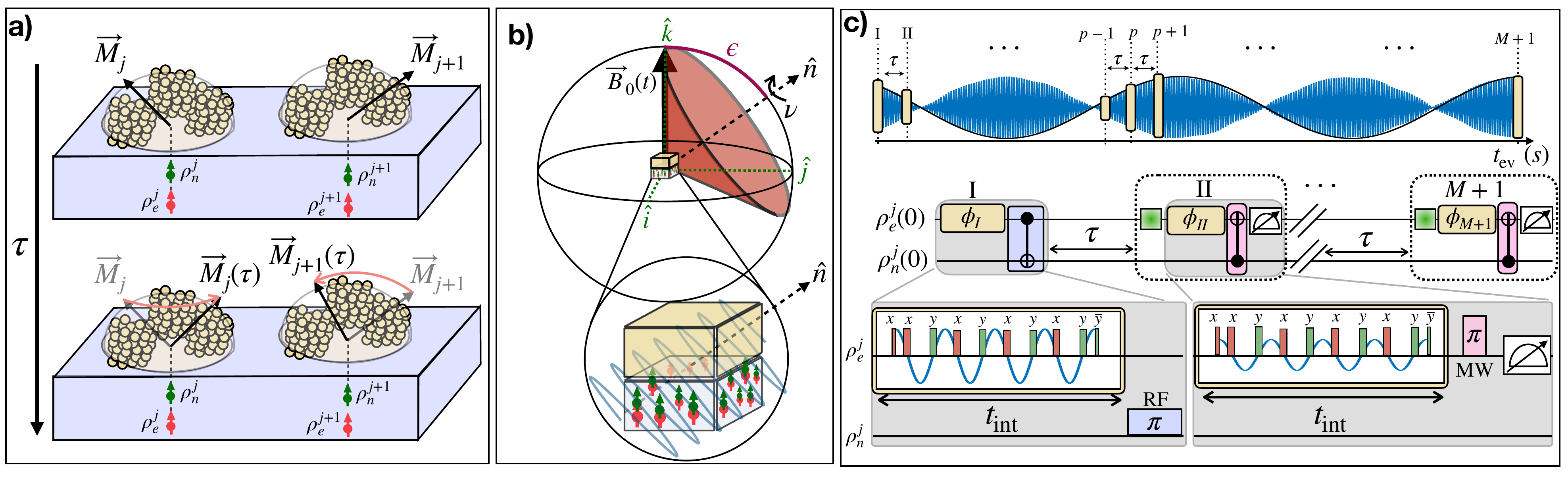}
\caption{\textbf{a)}~ \textbf{(Top)}~Each sensor comprising an electron and a memory qubit, probes a distinct region of the sample (for clarity, only the $j$-th and $(j+1)$-th units in the ensemble are depicted). These regions exhibit distinct initial magnetizations ($\vec{M_j}$ and $\vec{M_{j+1}} $) as a result of statistical polarisation. \textbf{(Bottom)} After a time $\tau$ each magnetization vector evolves owing to the combined action of \(\vec{B}_0(t)\) and  $H_c(t)$ in Eqs.~(\ref{eq:oscB}) and~(\ref{eq:driving}). \textbf{b)}~Trajectory of $\vec{B}_0(t)$ in Eq.~\eqref{eq:oscB}. This traces a cone, with aperture $\epsilon$ and central axis $\hat{n}$. The inset shows in more detail the diamond-sample block irradiated with RF and MWs. \textbf{c)}~ \textbf{(Top)} Signal emitted by the sample. The colored boxes correspond to the interrogation windows. \textbf{(Bottom)} Sequence over the $j$-th sensor. The initial states of the electron and nitrogen spins are $\rho_e^j(0)=\lvert 0\rangle_e^j\langle 0 \rvert_e^j$ and $\rho_n^j(0)=\lvert 0\rangle_n^j\langle 0 \rvert_n^j$. In region I, after an initial $\left(\pi/2\right)_x$ pulse, the electrons interrogate the sample's emitted signal (soft-yellow box), acquiring a phase $\phi_I^j$, which is then transformed back into populations by a $\left(\pi/2\right)_{-y}$ pulse and transferred to the nitrogen via a \(\mathrm{C}_{\mathrm{e}}\mathrm{NOT}_{\mathrm{n}}\) gate (soft-blue box). In regions $\{II,...,p,...,M+1\}$), the signal evolves for a time $\tau$ and is subsequently interrogated by the electron, which is  re-initialized (green boxes), and acquire a phase $\phi_{p}^j$. With a \(\mathrm{C}_{\mathrm{n}}\mathrm{NOT}_{\mathrm{e}}\) (soft-pink boxes), the initial phase \(\phi_{I}^j\) and $\phi_{p}^j$ are correlated and the population of the NV centers is measured.}
\label{fig:sequence}
\end{figure*}

In this work, we present a quantum sensing protocol for NV-center-based probing of solid-state nanoscale samples, which we identify as an ideal target for NV detection. Our method integrates three key ingredients: 
(1) In contrast to the conventional MAS picture, we here use a slow ($\sim$1~kHz) magnetic-field rotation~\cite{sakellariou2005nmr} to average out chemical shift anisotropy. This approach is equivalent to rotating the sample-NV assembly (see Appendices) but is simpler to depict schematically. This is synchronized with (2) RF irradiation to suppress dipolar couplings among target nuclei; and (3) a measurement scheme for coherent detection of the statistically polarized sample, enabling the characterization of key sample properties --specifically, isotropic chemical shifts.

The system,  comprising the sample and an ensemble of shallowly implanted NV centers is depicted in Fig.~\ref{fig:sequence}a (top), where each NV sensor (e.g., the $j$-th NV), consist of an electron and a nitrogen spin with quantum states $\rho_e^j$ and $\rho_n^j$. Since statistical polarization dominates, the initial sample magnetization above each sensor is different (i.e. $\vec{M}_j\neq \vec{M}_{j+1}$). Under the simultaneous influence of a rotating magnetic field and RF irradiation, the sample magnetization evolves as depicted in Fig.~\ref{fig:sequence} a) (bottom). We demonstrate that this evolution encodes only the isotropic part of the chemical shifts while mitigating the strong dipole-dipole interactions. Subsequently, this evolution is probed by each NV sensor.

We first examine the sample evolution under the proposed protocol and subsequently introduce a readout method for NV-based interrogation.
For a solid-state system of $N$ homonuclear spins subjected to an external magnetic field $\vec{B}_0 = B_0 \hat{b}_0$ in an arbitrary orientation, the Hamiltonian reads
\begin{align}
H=\sum_k^N\left[\gamma\:\vec{I}^k\cdot\left(\mathds{1}+\boldsymbol{\sigma^k}\right)\cdot\vec{B}_0\right]+H_{dd}+H_c(t),\label{eq:H}
\end{align}
where $\vec{I}^k$, $\gamma$, and $\boldsymbol{\sigma^k}$ denote the spin operator, the gyromagnetic ratio, and the chemical shift tensor of the $k$-th spin. The term $H_{dd}$ accounts for internuclear dipolar interactions, while $H_c(t)$ represents RF control fields.  The structural and chemical environment of the sample is defined by the set of tensors $\{\boldsymbol{\sigma}^k\}_{k=1}^N$. For the $k$-th spin, the eigenvectors of $\boldsymbol{\sigma}^k$ define the principal axes system (PAS), denoted as $\{X_k, Y_k, Z_k\}$, while the corresponding eigenvalues $\{\sigma_{X_k}, \sigma_{Y_k}, \sigma_{Z_k}\}$ are the chemical shifts~\cite{levitt2008spin}.  

Under the secular approximation, $\gamma\lvert\vec{B_0}\rvert~\gg~\gamma\lvert\boldsymbol{\sigma^k}\cdot\vec{B}_0\rvert$, the  contribution including  $\boldsymbol{\sigma^k}$ reduces to $$\sigma^k(\vec{\theta})\;I_{\hat{b}_0}^k~=~\left[\cos^2{\left(\theta^k_x\right)}\sigma_{X_k}~+~\cos^2{\left(\theta^k_y\right)}\sigma_{Y_k}~+~\cos^2{\left(\sigma^k_z\right)}\sigma_{Z_k}\right]I_{\hat{b}_0}^k,$$ with $\theta^k_i$ the angles between each principal axis of the $k$th spin and $\hat{b}_0$~\cite{levitt2008spin}. A schematic illustration of the scenario described above, including all the axes and angles, is shown in Fig.~\ref{fig:geometric} a) in the Appendices. Within this context, orientation-dependent peak shifts (governed by $\theta^k_i$) degrade both resolution and sensitivity. Furthermore, strong dipolar interactions ($\sim$ tens of kHz) obscure the spectral features. In liquid-state samples, these challenges are inherently mitigated by rapid molecular diffusion and tumbling. This effectively cancels dipolar couplings, $\overline{H_{dd}^{k,l}}\rightarrow 0$, and reduces each  chemical shift to a single, orientation-independent term $\overline{\sigma^k(\vec{\theta})}=\frac{\sigma_{X_k}+\sigma_{Y_k}+\sigma_{Z_k}}{3}\equiv\sigma_{\rm iso}^k$ where overline indicates time-averaging.

In the solid state, however, the absence of  molecular motion prevents the averaging of dipole-dipole interactions and chemical shift anisotropy. To circumvent this, firstly, our protocol introduces a slowly rotating magnetic field of the form
\begin{align}
\vec{B}_0(t)=B_0\left[\sin{\epsilon}\sin{\left(\nu t\right)\;\hat{u}}-\sin{\epsilon}\cos{\left(\nu t\right)}\;\hat{v}+\cos{\epsilon}\;\hat{n}\right],\label{eq:oscB}
\end{align}
with $\{\hat{u},\hat{v},\hat{n}\}=\{\hat{i},\cos{\epsilon}\;\hat{j}-\sin{\epsilon}\;\hat{k},\;\sin{\epsilon}\;\hat{j}+\cos{\epsilon}\;\hat{k}\}$.
As shown in Fig.~\ref{fig:sequence} b), the vector $\vec{B}_0(t)$ traces a conical trajectory of aperture $\epsilon$ about the axis $\hat{n}$ at frequency $\nu\sim 1$ kHz.

Secondly, we apply a tailored RF  field, described by 
\begin{align}
        H_c(t)=2\Omega\cos{\left(\omega t+\alpha\right)}\sum_{k=1}^N\left[I_{u}^k\;\cos{\left(\nu t\right)}+I_{v}^k\;\sin{\left(\nu t\right)}\right]
    \label{eq:driving}
\end{align}
where \(\Omega\), \(\omega\), and \(\alpha\) denote the amplitude, frequency, and phase offset, respectively.  Note that the driving direction is modulated at the same frequency $\nu$. Dipolar decoupling is achieved by setting the carrier frequency to $\omega = \gamma \lvert\vec{B}_0(t)\rvert - \Delta$, with an off-resonance condition $\Delta = \Omega/\sqrt{2}$. To ensure protocol stability, we alternate the sign of $\Omega$ to emulate a frequency-switched Lee-Goldburg (fsLG) scheme~\cite{bielecki1989frequency,halse2013improved}. This approach provides resilience against various error sources, as demonstrated in the numerical results below.

As shown in Appendix~\ref{app:dyn_ensemble}, after a period $\tau~=~1/\nu$ of $\vec{B}_0(t)$, and under the action of $H_c(t)$ the dynamics of the sample is governed by
\begin{equation}
 H_{II}\approx\frac{1}{\sqrt{3}}\sum_{k=1}^N\left[\gamma\lvert\vec{B_0}\rvert\delta_{\rm iso}^{(k)}+\frac{\left(\Omega\right)^2}{4\omega}\right]I_{n}^k,\label{eq:traj}
\end{equation}
where the first term recovers the isotropic chemical shift, and the second contains a Bloch-Siegert shift (see Appendix~\ref{app:BSshift}). 
Equation~(\ref{eq:traj}) indicates that the combination of $\vec{B}_0(t)$, and $H_c(t)$ suppresses chemical shift anisotropy and dipolar interactions, isolating the isotropic chemical shifts $\delta_{\rm iso}^{(k)}$ as a distinctive sample fingerprint. As detailed below, our numerical analysis demonstrates the robustness of this sensing protocol against experimental non-idealities, including field misalignments and stochastic noise in $\Omega$.

Based on Eq.~\eqref{eq:traj}, we define an interrogation window $t \in [0, t_{\text{int}}]$ spanning an integer number $p$ of magnetic-field periods ($t_{\text{ev}} = p\tau$). Within this window, the $j$-th sensor detects a signal (see Appendix~\ref{app:signal}) of the form:
\begin{align}
	s^j(t)\approx \bigg[A^j \sum_{k=1}^N\langle I_{u}^k\rangle\left(t_{\rm ev}\right) \bigg]\cos\!\left(\Omega t\right),\label{eq:signal}
\end{align}
During the detection window, the signal $s^j(t)$ oscillates at the RF Rabi frequency $\Omega$, with an amplitude  
\(A^j\) which is slowly modulated by the term $\langle I_{u}^k\rangle\left(t_{\rm ev}\right)=
\cos\!\left[\frac{1}{\sqrt{3}}\Big(\delta_{\rm iso}^{(k)}+\frac{\Omega^{2}}{6\omega}\Big)t_{\mathrm{ev}}+\xi^{j}\right]$
where $\xi^j$ is an initial phase (which generally differs from one sensor to another).

The sample's emitted signal for our numerical example (see later) is illustrated in Fig.~\ref{fig:sequence} c) (top) by a solid blue line. The black solid line denotes the slow amplitude modulation, $\langle I_{u}^k\rangle\left(t_{\rm ev}\right)$, that encodes \(\delta_{\rm iso}^{(k)}\). Shaded yellow regions denote interrogation windows at $t_{\text{ev}} = p\tau$ for distinct values of $p$.

Phase acquisition by the NV sensors occurs during the interrogation windows, i.e. once every period $\tau$. The $j$th NV-signal Hamiltonian, comprising electron \(\vec{S}^j\) and nitrogen  \(\vec{P}^j\) spins, and  $s^j(t)$ is
\begin{align}
	H_{\rm sensor}^j = \gamma_e s^j(t)\;S_z^j + \vec{S}^j\cdot\hat{\textbf{A}}\cdot\vec{P}^j + H_{\rm MW} + H_{\rm RF}^{N},\label{eq:sensor}
\end{align}
where $\gamma_e$ is the electron gyromagnetic ratio, $\hat{\mathbf{A}}$ is the dipolar tensor between the electron and nitrogen, and $H_{\rm MW}$ and $H_{\rm RF}^{N}$ are  controls (MW and RF), see Fig.~\ref{fig:sequence}c (bottom). 

The initial  NV state is $\rho_e^j(0)\otimes \rho_n^j(0)$ with the polarized components $\rho_e^j(0)=\lvert 0\rangle_e^j\langle 0 \rvert_e^j$ and $\rho_n^j(0)=\lvert 0\rangle_n^j\langle 0 \rvert_n^j$~\cite{spohn2025quantum}. During each interrogation window, the $j$th sensor probes the signal  $s^j(t)$ by first applying a $\left(\pi/2\right)_x$ pulse, followed by a dynamical decoupling sequence (e.g., XY8) with the interpulse spacing set on resonance to $1/\left(2\Omega\right)$. As described in Appendix~\ref{app:sensor}, at the $p$th interrogation window the accumulated phase is
\begin{align}
\phi_j^{(p)}\propto A^j\cos{\left(\xi_j+\frac{1}{\sqrt{3}}\left[\delta_{\rm iso}+\frac{\Omega^2}{4\omega}\right]\;p\tau\right)}.
\end{align}
Recall that, since statistical polarization dominates, \(A^j\) and \(\xi^j\) differ among sensors, see Fig.~\ref{fig:sequence}a). Therefore, a direct measurement of $\phi_j^{(p)}$ produces response that cancels when averaging over the ensemble. 
The  flexibility of our method allows us to incorporate advanced correlation spectroscopy techniques, in particular \textit{multi-point correlation spectroscopy}~\cite{spohn2025quantum}, to mitigate these effects as illustrated in Fig.~\ref{fig:sequence}c.  In particular, At $p=0$, region I in Fig.~\ref{fig:sequence}~c), the acquired phase $\phi_j^{(0)}$ is converted into populations by a $\left(\pi/2\right)_{-y}$ pulse and stored in the nitrogen spin via a $\mathrm{C}_{\mathrm{e}}\mathrm{NOT}_{\mathrm{n}}$ gate (via a weak RF pulse). For $p>0$ (regions $\{{\rm II,\dots,M+1}\}$ in Fig.~\ref{fig:sequence}~c)) the corresponding phase $\phi_j^{(p)}$ is correlated with $\phi_j^{(0)}$ using a $\mathrm{C}_{\mathrm{n}}\mathrm{NOT}_{\mathrm{e}}$ gate and, finally,  the electron spin state is read out. Appendix~\ref{app:sensor} demonstrates that the final expression becomes
\begin{align}
    \langle \overline{S_z}\rangle=g\sum_{i=1}^lN^{(i)}\cos{\left(\frac{1}{\sqrt{3}}\left[\delta_{\rm iso}^{(i)}+\frac{\Omega^2}{4\omega}\right]p\tau\right)},\label{eq:result}
\end{align}
where $\langle \overline{S_z}\rangle$ denotes averaging over the NV ensemble. Other relevant parameters are $g=\left(\frac{2\gamma_e\gamma_n t_{\rm int}}{\pi}\frac{\mu_0h}{8\pi}\frac{1}{2}\right)^2\frac{F_z^{(2)}}{3}$ and $\displaystyle F_z^{(2)}$ a geometric factor that, typically, comprises a semi-bubble with a radius on the order of the NV depth (see ~\cite{glenn2018high} and Appendix~\ref{app:signal}). In addition, detection volume was assumed containing $N$ nuclei of the same species, divided into $l$ groups with different chemical shift tensors, such that $\displaystyle\sum_{i=1}^l N^{(i)} = N$.

\begin{figure}[t!!]
\includegraphics[width=1 \linewidth]{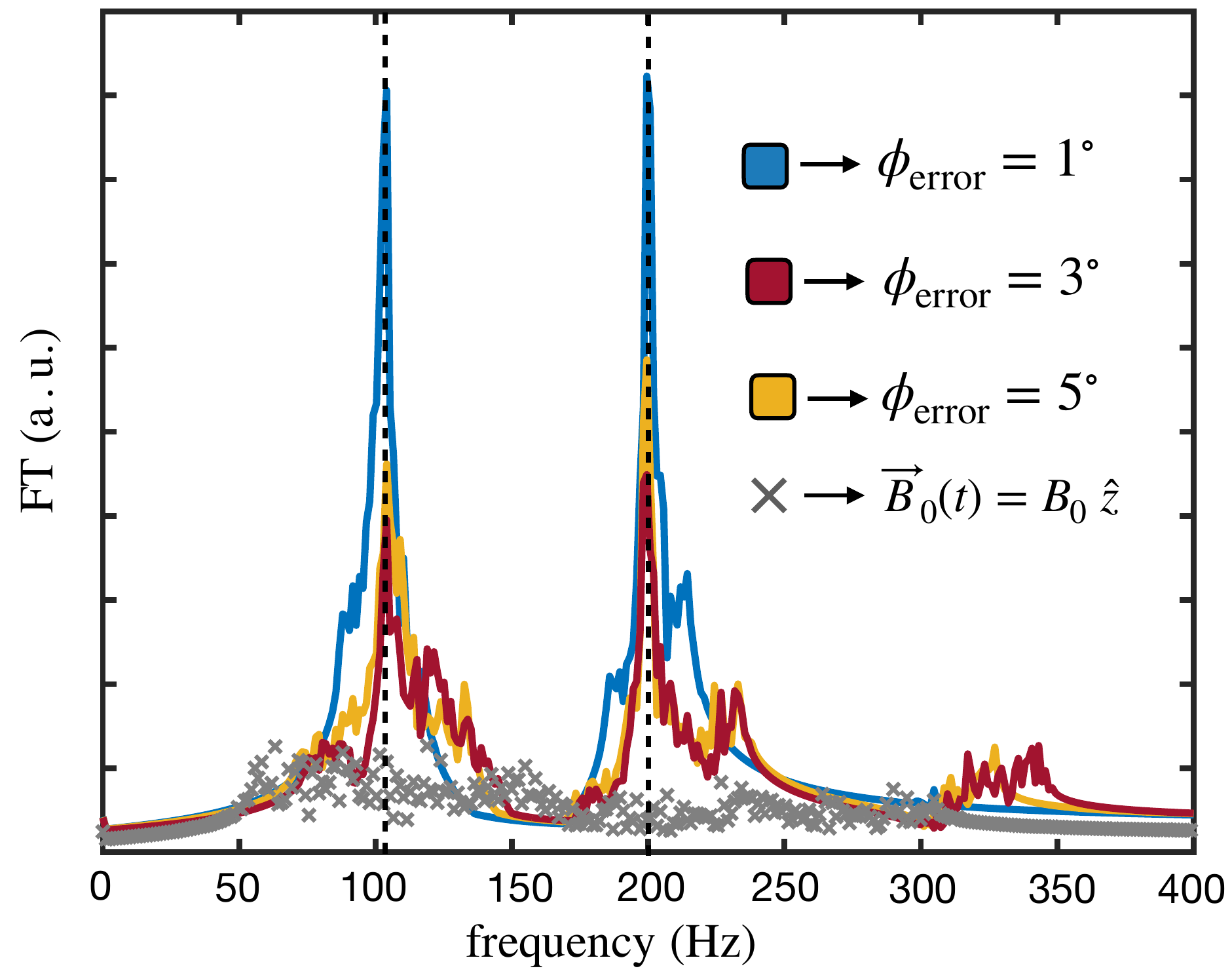}
\caption{Simulated spectrum for a sample with two magnetically non-equivalent protons. The solid colored lines represent the spectra obtained using our protocol under various misalignment errors, while gray crosses depict the typical powder spectrum (for a non-rotating magnetic field) for comparison. Dashed vertical lines indicate the theoretical resonances predicted by our framework. The results demonstrate the protocol's robustness against both increasing misalignments and stochastic amplitude noise ($0.25\%$ fluctuations with a $1$~ms correlation time).}
\label{fig:results_hf}
\end{figure}

Finally, we validate our theoretical framework through numerical simulations of the proposed protocol. While the alternative implementation involving the rotation of the NV-sample block is detailed in Appendix~\ref{app:variation}, both approaches yield equivalent results, differing only in their resilience to specific experimental errors.
To validate the approximations made in our theoretical derivation, we perform a full numerical simulation of the protocol. In particular, the sample evolution is integrated directly from the complete Hamiltonian in Eq.~\eqref{eq:H} without further simplifications. We take as a figure of merit a sample containing pairs of two distinct hydrogen nuclei with random orientations. The chemical-shift principal values are $\delta_{X_P^1}=352$ Hz, $\delta_{Y_P^1}=22$ Hz, $\delta_{Z_P^1}=456$ Hz, $\delta_{X_P^2}=221$ Hz, $\delta_{Y_P^2}=27$ Hz, and $\delta_{Z_P^2}=74$ Hz, leading to $\delta_{\rm iso}^{(1)}=277$ Hz and $\delta_{\rm iso}^{(2)}~=~107$~Hz. We consider an NV depth of $\sim 5\,\mathrm{nm}$, which leads to a semispherical detection volume of approximately $125\,\mathrm{nm}^3$.  The magnetic-field rotation period is $1/\nu=1\,\mathrm{ms}$, and the nitrogen-spin thermalisation time is $T_1=1\,\mathrm{s}$. The internuclear distance is $0.25\,\mathrm{nm}$, corresponding to a maximum dipolar interaction of $\sim 14.9\,\mathrm{kHz}$. The external magnetic field is $B_0=2\,\mathrm{T}$, corresponding to a Larmor frequency of $\omega=84\,\mathrm{MHz}$. The driving amplitude is $\Omega\approx154\,\mathrm{kHz}$, which gives a Bloch--Siegert shift of $\frac{1}{\sqrt{3}}\Omega^2/(4\omega)=40.2\,\mathrm{Hz}$. 

The driving was simulated with stochastic amplitude errors of $0.25\%$ and a correlation time of $1\,\mathrm{ms}$. Also various misalignment angles were assumed that modifies Eq.~\eqref{eq:driving} to
\begin{align}
H_c(t)=2\Omega\cos{\left(\omega t+\alpha\right)}\sum_{k=1}^N\left[a\,I_{\hat{u}}^k\;\sin{\left(\nu t\right)}+b\,I_{\hat{v}}^k\;\cos{\left(\nu t\right)}+c\,I_{\hat{n}}^k\right],\nonumber
\end{align}
where \[
a\hat{u}+b\hat{v}+c\hat{n}=\cos(\phi_{\rm error})\,\mathbf{v}_0+\sin(\phi_{\rm error})\,\mathbf{u}.
\]
Here, \(
\mathbf{v}_0=\left(\hat{u}+\hat{v}\right)
\) is the ideal direction and \(\mathbf{u}\) a random orthogonal direction that models misalignment. Ideally, $\phi_{\rm err}=0 \rightarrow a=1,\ b=1,\ c=0$, recovering Eq.~\eqref{eq:driving}. For completeness, Appendix~\ref{app:misa} discusses an alternative RF driving scheme that, while potentially offering a simpler experimental implementation, exhibits a larger susceptibility to misalignment.

In Fig.~\ref{fig:results_hf}, we show the simulated spectrum obtained by applying our method. All cases include stochastic amplitude errors and the different colored solid lines correspond to increasing misalignment values ($\phi_{\rm error}=1^{\circ}$, $3^{\circ}$, and $5^{\circ}$). Our method remains robust against these imperfections, recovering the isotropic contribution in all cases. For completeness, gray crosses correspond to the spectrum obtained when the external magnetic field is not rotating, recovering the usual powder spectrum. This significantly reduces sensitivity and resolution (see Fig.~\ref{fig:results_hf}). The resonances predicted by our theory --specifically by Eq.~\eqref{eq:result}-- are indicated by dashed vertical lines in Fig.~\ref{fig:results_hf}. For this configuration, the predicted values of $\frac{1}{\sqrt{3}}\left(\delta_{\rm iso}^{(1)}+\frac{\Omega^2}{4\omega}\right)\approx200\ {\rm Hz}$  and $\frac{1}{\sqrt{3}}\left(\delta_{\rm iso}^{(2)}+\frac{\Omega^2}{4\omega}\right)\approx102\ {\rm Hz}$ are in excellent agreement with the numerical results.

In summary, our results showcase the potential of NV detection for nanoscale solid-state samples. By integrating homogeneous RF decoupling with the suppression of chemical shift anisotropy, our ensemble-based approach achieves  chemical-shift resolution. The proposed protocol integrates magnetic field rotation and memory-qubit storage to extract coherent information from statistically polarized solid-state samples. Numerical simulations confirm the robustness of this approach against experimental non-idealities, demonstrating a perfect match with our theoretical framework

Authors acknowledge support by the European Union's Horizon Europe research and innovation programme under Grant Agreement No. 101135742 (QUENCH). P.A.B. acknowledges support from
UPV/EHU Ph.D. Grant PIF 23/275. I.I.Z. acknowledges support from UPV/EHU Ph.D. Grant PIF 23/246. J. C. acknowledges the Agencia Estatal de Investigaci\'{o}n via the Modelizado, Optimizaci\'{o}n, y Esquemas de Magnetometria en Centros de Color project PID2024-161371NB-C22, and the Basque Government under Grant No. IT1470-22.  
D.B.B acknowledges the Deutsche Forschungsgemeinschaft (DFG, German Research Foundation)--412351169 within the Emmy Noether program and the support by the DFG under Germany's Excellence Strategy--EXC 2089/1--390776260 and the EXC-2111 390814868.

\newpage
\onecolumngrid
\appendix
\counterwithin{figure}{section}

\newpage
\section{Theory of the ensemble}\label{app:dyn_ensemble}
The Hamiltonian is given by Eq.~\eqref{eq:H}

\begin{align}
H=\sum_k^N\left[\gamma^k\:\vec{I}^k\cdot\left(\mathds{1}+\boldsymbol{\sigma^{(k)}}\right)\cdot\vec{B}_0(t)\right]+\sum_{k<l}^NH_{dd}^{k,l}+H_{\rm c}(t).\label{eq:appA:H}
\end{align}
Each CS tensor \(\boldsymbol{\sigma^{(k)}}\) has three preferred, mutually perpendicular directions along which the tensor is diagonal. These directions define a reference frame commonly called the principal-axis system (PAS). Although the CS tensor can be expressed in any coordinate frame, it is convenient to write it as a rotation from the PAS to the observers frame (i.e. LAB frame):

\begin{align}
\boldsymbol{\sigma^{(k)}}=R
\begin{pmatrix}
    \sigma_{X_k} & 0 & 0\\
    0 & \sigma_{Y_k} & 0\\
    0 & 0 & \sigma_{Z_k}
\end{pmatrix}
R^{-1}=R\sigma^{(k)}_{\rm PAS}R^{-1}.\label{eq:app:cs_pas}
\end{align}
$R$ is the rotation from the PAS to the LAB frame.

The goal is to average the anisotropy arising from the CS tensors $\boldsymbol{\sigma^{(k)}}$ and obtain the dynamics of the isotropic part (i.e. the mean of the three principal values). To achieve so, we combine a slowly rotating magnetic field,
\begin{align}
\vec{B}_0(t)=B_0\left[\sin{\epsilon\cos\left(\nu t\right)\:\hat{u}+\sin{\epsilon}\sin{\left(\nu t\right)\:\hat{v}+\cos{\epsilon}}}\:\hat{n}\right].\nonumber
\end{align} 
Also, the homonuclear dipole-dipole interaction $H_{ dd}^{k,l}$ is averaged out by means of continuous decoupling RF radiation fulfilling the LG condition (see further below for explicit conditions)
\begin{align}
        H_c(t)=2\Omega\cos{\left(\omega t+\alpha\right)}\sum_{k=1}^N\left[I_{\hat{u}}^k\;\cos{\left(\nu t\right)}+I_{\hat{v}}^k\;\sin{\left(\nu t\right)}\right], \label{eq:app:LG}
\end{align}
Note that the dominant term of the Hamiltonian \eqref{eq:appA:H}, the Zeeman interaction
\(\displaystyle \sum_{k=1}^N \gamma^k\,\vec{I}^k\cdot\vec{B}_0(t)\), has a slow time-dependency. However, the rotation frequency \(\nu\) is assumed to be much smaller than the Zeeman splitting \(\gamma^k B_0\), so the field magnitude is effectively constant during the periods $\frac{1}{\Omega}$ and $\frac{1}{\omega}$. We define the instantaneous quantisation axis for spin \(k\) as
\[
I_{S(t)}^k \equiv \frac{\vec{I}^k\cdot\vec{B}_0(t)}{B_0}=\sin{\epsilon}\left[\sin{\left(\nu t\right)}\:I_{\hat{u}}^k-\cos{\left(\nu t\right)}\:I_{\hat{v}}^k\right]+\cos{\epsilon}\:I_{\hat{n}}^k.
\]
These instantaneous axes form a time-dependent reference frame that we call the MAS frame:
\begin{align}
I_{M(t)}^k&=\cos{\left(\nu t\right)}\:I_{\hat{u}}^k+\sin{\left(\nu t\right)}\:I_{\hat{v}}^k,\nonumber\\[8pt]
I_{A(t)}^k&=-\cos{\epsilon}\left[\sin{\left(\nu t\right)}\:I_{\hat{u}}^k-\cos{\left(\nu t\right)}\:I_{\hat{v}}^k\right]+\sin{\epsilon}\:I_{\hat{n}}^k,\nonumber\\[8pt]
I_{S(t)}^k&=\sin{\epsilon}\left[\sin{\left(\nu t\right)}\:I_{\hat{u}}^k-\cos{\left(\nu t\right)}\:I_{\hat{v}}^k\right]+\cos{\epsilon}\:I_{\hat{n}}^k.
\label{eq:app:osc_axes}
\end{align}
Note that with the new axes, the driving \eqref{eq:app:LG} reads
\begin{align}
        H_{c}(t)&=\sum_{k=1}^N\left[2\;\Omega^kI_{M(t)}^k\ \cos\left(\omega^kt+\alpha\right)\right],\nonumber
\end{align}

In terms of the new set of axes \eqref{eq:app:osc_axes}, and rewriting the CS tensor in terms of the PAS \eqref{eq:app:cs_pas}, the Hamiltonian \eqref{eq:appA:H} is
\begin{align}
    H(t)=B_0\sum_k^N\left[\gamma^k\ \left(I_{S(t)}^k+\vec{I^k}\cdot R_k\sigma^{(k)}_{\rm PAS}R^{-1}_k\cdot\hat{B}_0\right)\right]+\sum_{k<l}^NH_{dd}^{k,l}+H_{\rm c}(t),\label{eq:appA:inH}
\end{align}

Assuming a detuned driving, $\omega^k=\gamma^k\lvert\vec{B}_0\rvert-\Delta^k$, in the IP with respect to $H_0=\displaystyle\sum_k^N\left[\left(B_0\gamma^k-\Delta^k\right)\ I_{S(t)}^k\right]$, and invoking the RWA, one obtains
\begin{align}
    H_I(t)&=\sum_k^N\left[\cos^2{\left(\theta_x^k\right)}\delta_{X_k}+\cos^2{\left(\theta_y^k\right)}\delta_{Y_k}+\cos^2{\left(\theta_z^k\right)}\delta_{Z_k}\right]I_{S(t)}^k+\sum_{k<l}^N\tilde{H}_{dd}^{k,l}\nonumber\\[6pt]
    &+\sum_k^N\left[\Delta^k\ I_{S(t)}^k+\Omega^k\;I_{\alpha(t)}^k\right],\nonumber
\end{align}
where we defined $\delta_i\equiv B_0\gamma^{(k)}\sigma_i$, neglected the fast oscillatory terms and use $I_{\alpha(t)}^k\equiv\sin{\alpha\ I_{A(t)}^k}+\cos{\alpha}\ I_{M(t)}^k$.  Here, $\theta_x^k$,  $\theta_y^k$ and  $\theta_z^k$ appear in the explicit form of $R$ in Eq.~\eqref{eq:app:cs_pas}, which is the rotation matrix from the PAS to the MAS frame. Thus, $\theta_x^k$,  $\theta_y^k$ and  $\theta_z^k$ are the angles between the three principal axes of the k-th spin with the direction of the magnetic field, $\hat{S}$.
To achieve dipolar decoupling, we choose the frequency switched Lee-Goldburg  (fsLG) method, thus one must set $\Delta^k=\pm\frac{\;\Omega^k}{\sqrt{2}}$. Thus, in what follows, the term $H_{dd}^{k,l}$ is assumed to be suppressed.
\begin{figure*}[]
\includegraphics[width=0.6 \linewidth]{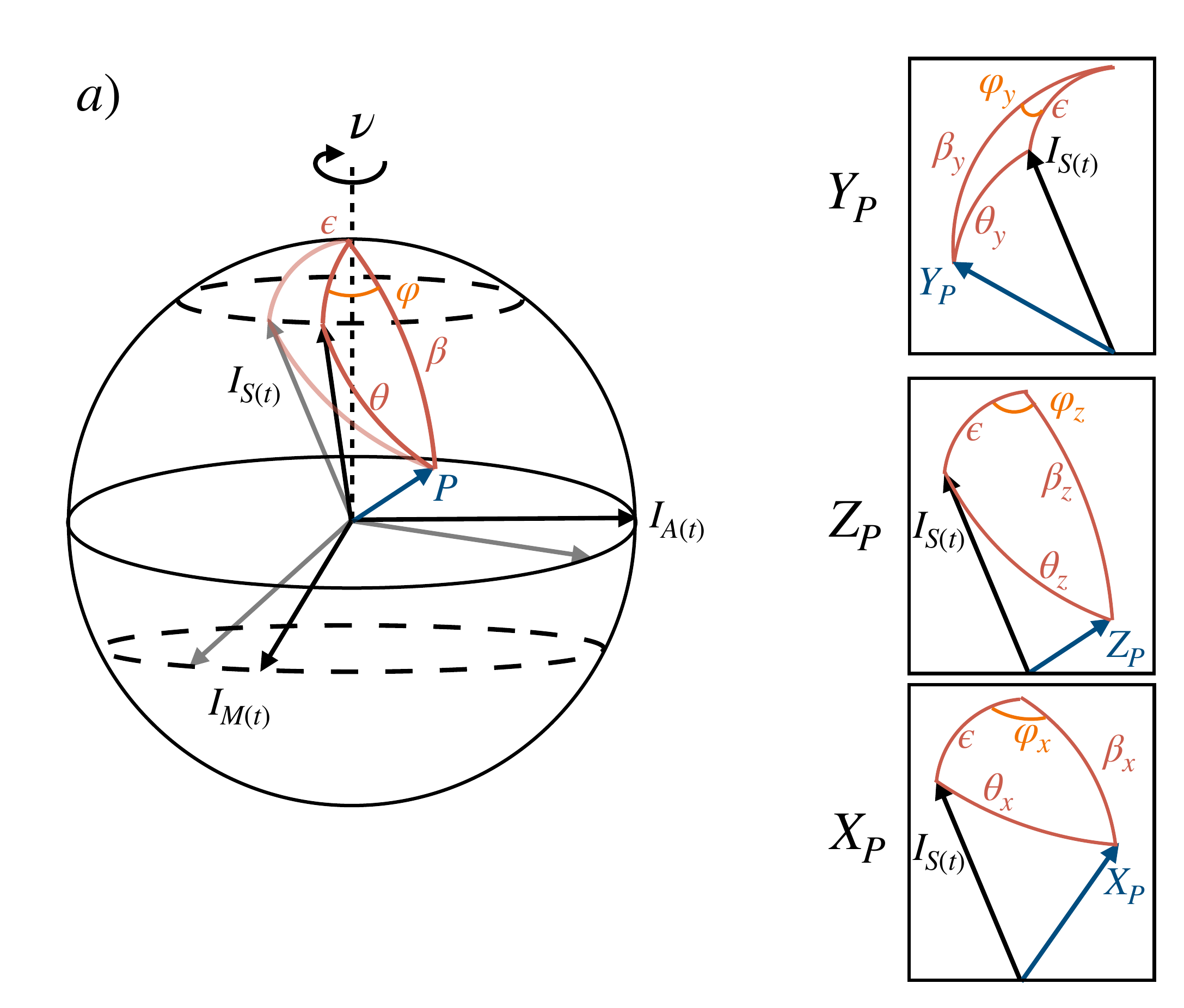}
\caption{Schematic representation of the new reference frame defined by the tilted oscillation of the magnetic field and the PA of a given molecule. a) The initial axes $\{x_0,y_0,z_0\}$ are depicted as solid black vectors. These are defined by the first magnetic field position ($z_0$) with respect to the fixed LAB frame. The angle $\epsilon=\arccos{\frac{1}{\sqrt{3}}}$ is the angle between the magnetic field and the fixed vertical LAB axis. The PA of a given molecule are depicted $\{x_P,y_P,z_P\}$ as solid blue vectors. }
\label{fig:geometric}
\end{figure*}

In the I.P. with respect to the driving, assuming that the effective Rabi frequency, $\Tilde{\Omega}^k=\sqrt{\left(\;\Omega^k\right)^2+\left(\Delta^k\right)^2}$, is greater than any other interaction, one gets
\begin{align}
    H_{II}(t)&=\sum_k^N\left\{\left[\cos^2{\left(\theta_x^k\right)}\delta_{X_k}+\cos^2{\left(\theta_y^k\right)}\delta_{Y_k}+\cos^2{\left(\theta_z^k\right)}\delta_{Z_k}\right]\left[\frac{I_{S(t)}^k+\sqrt{2}\ I_{\alpha(t)}^k}{3}\right]\right\},\label{eq:app:H2}
\end{align}
It is useful to write $\theta^i_k$ in terms the magnetic field tilting angle $\epsilon$, the oscillation frequency $\nu$ and two extra angles $\beta_i^k$ and $\varphi_i^k$ that depend on the sample's orientation. Fig.~\ref{fig:geometric} illustrates this configuration. The magnetic field direction oscillates about the vertical axis (i.e. $\hat{n}=\hat{z}$ is chosen for clarity) with frequency \(\nu\). Each molecule's principal axis is fixed, tracing an angle \(\theta_i^k\) with the instantaneous field direction $\hat{S}$. Thus, $\theta_i^k$ are indeed time dependent. 
The arcs corresponding to \(\theta\), \(\beta\) and \(\epsilon\) form a spherical triangle. Applying the spherical law of cosines to that triangle, one finds
\begin{align}
\cos^2{\theta_i(t)}&=\cos^2{\epsilon}\cos^2{\beta_i}+\sin^2{\epsilon}\sin^2{\beta_i}\cos^2{\varphi_i(t)}+2\cos{\epsilon}\cos{\beta_i}\sin{\epsilon}\sin{\beta_i}\cos{\varphi_i(t)}\nonumber\\[8pt]
    &=\frac{1}{2}\left[\left(3\cos^2{\epsilon}-1\right)\cos^2{\beta_i}+\sin^2{\epsilon}\right]+\frac{\sin^2{\epsilon}\sin^2{\beta_i}\cos\{2\varphi_i(t)\}}{2}+\cos{\epsilon}\sin{\epsilon}\cos{\beta_i}\sin{\beta_i}\cos{\varphi_i(t)},\nonumber
\end{align}
where $\varphi_i(t)=\varphi_i^0+\nu t$. If $\epsilon=\arccos{\left(\frac{1}{\sqrt{3}}\right)}$, this expression reduces to 
\begin{align}
    \cos^2{\theta_i(t)}=\frac{1}{3}+\frac{1}{3}\sin^2{\beta_i}\cos\{2\varphi_i^0+2\nu t\}+\frac{\sqrt{2}}{3}\cos{\beta_i}\sin{\beta_i}\cos\{\varphi_i^0+\nu t\},\nonumber
\end{align}
In addition, if $\alpha=\pi/2$ then $$\frac{I_{S(t)}^k+\sqrt{2}\ I_{\alpha(t)}^k}{3}=\frac{I_{\hat{n}}^k}{\sqrt{3}}.$$
The Hamiltonian \eqref{eq:app:H2} is then
\begin{align}
    H_{II}(t)=\frac{1}{\sqrt{3}}\sum_k^N\left\{\sum_{i=x,y,z}\frac{\delta_{i_k}}{3}\left[1+A_i^k(t)+C_i^k(t)\right]I_{\hat{n}}^k\right\},\nonumber
\end{align}
where we defined 
\begin{align}
A_i^k(t)&=\sin^2{\beta_i^k}\cos\{2\varphi_i^{k,0}+2\nu t\},\nonumber\\[6pt]
C_i^k(t)&=\sqrt{2}\cos{\beta_i^k}\sin{\beta_i^k}\cos\{\varphi_i^{k,0}+\nu t\}\nonumber
\end{align}

We study the effective Hamiltonian using average Hamiltonian theory. Since the Hamiltonian commutes at all times, the first order in of the AHT expansion is the only contribution. 
\begin{align}
    H_{II}^{(1)}=\frac{1}{T}\int_0^T H_{II}(s)ds=\sum_k^N\left[\frac{\delta_{\rm iso}^{(k)}}{\sqrt{3}}I_{\hat{n}}^k\right]+\frac{1}{T}\sum_k^N\left[\frac{A_i^k(T)\sin{\left(\nu T\right)}+2C_i^k(T)\sin{\left(\nu T/2\right)}}{2\nu}\right]I_{\hat{n}}^k.\nonumber
\end{align}
The first term is the isotropic contribution, $\delta^{(k)}=\frac{\delta_{X_k}+\delta_{Y_k}+\delta_{Z_k}}{3}$, which is our target. The second term is undesired. Crucially, this term is exactly zero if $T=\frac{1}{\left(2\pi\right)\times\nu}$. Thus, if the sample is probed once every magnetic field period $T=1/\nu$, it will not be observed. In addition, such contribution is bounded: while the isotropic part accumulates over the evolution (growing proportionally with time), the second term does not, so its relative importance decreases at long times. Thus, even if probing the sample before one full period of the magnetic field, the contribution of such term is negligible compared to the isotropic chemical shift. Increasing \(\nu\) further reduces the relative weight of the undesired term.

Then, we can safely assume
\begin{align}
    H_{II}^{(1)}\approx\sum_{k=1}^N\left[\frac{\delta_{\rm iso}^{(k)}}{\sqrt{3}}I_{\hat{n}}^k\right].\label{eq:appA:dynamics}
\end{align}

Since this Hamiltonian is derived in the interaction picture with respect to $H_0=\sum_{k=1}^N\left[\left(B_0\gamma^k-\Delta^k\right) I_{S(t)}^k\right]$,
but we want it to hold in the LAB frame, one must choose \(\Omega\) to avoid spurious frequency artifacts. This is achieved by imposing
\begin{align}
\sqrt{\left(\Omega^k\right)^2+\left(\Delta^k\right)^2}=\frac{\omega^k}{p}
\quad \Longrightarrow \quad
\Omega^k=\sqrt{\frac{2}{3}}\frac{\omega^k}{p},
\nonumber
\end{align}
where $p$ is a positive integer number.
\newpage

\section{Bloch-siegert shift}\label{app:BSshift}
In scenarios with strong irradiation of the sample, one typically observes a shift in the spins' resonance frequency. This shift is attributed to the contribution of the counter-rotating terms associated with the driving field, which are usually neglected within the rotating wave approximation (RWA). In this section, we compute the effect of these counter-rotating terms on the dynamics of our protocol. The initial Hamiltonian is Eq.~\eqref{eq:appA:inH}, (assuming a correct cancelation of the dipolar interaction)
\begin{align}
    H(t)=\sum_k^NB_0\gamma^k\left[\ I_{S(t)}^k+\vec{I^k}\cdot R_k\sigma^k_{\rm PAS}R^{-1}_k\cdot\hat{B}_0\right]+\sum_{k=1}^N\left[2\;\Omega^kI_{M(t)}^k\ \cos\left(\omega^kt+\alpha\right)\right].\nonumber
\end{align}
With $\omega^k=B_0\gamma^k-\Delta^k$ and assuming that $I_S(t)$ does not change during the driving time-scale $\frac{1}{\Omega^k}$, moving to the I.P. with respect to $H_0=w^kI_S$ while keeping the counter-rotating terms, one gets (recall that $\alpha=\pi/2$)
\begin{align}
H_I(t)=\sum_k^N&\left[A_{\rm CS}^kI_S^k+\Delta^k I_S^k+\Omega^kI_A^k+\Omega^k\left(I_A\cos{\left(2\omega^kt\right)}-I_M\sin{\left(2\omega^kt\right)}\right)\right].\nonumber
\end{align}
Here, we use $A_{\rm CS}^k=B_0\gamma^k\left[\cos^2{\left(\theta_k^x\right)}\sigma_{X_k}+\cos^2{\left(\theta_k^y\right)}\sigma_{Y_k}+\cos^2{\left(\theta_k^z\right)}\sigma_{Z_k}\right]$.

Now, let us compute the first and second order Hamiltonians using average Hamiltonian theory. To avoid confusion, we are going to explicitly write all the $(2\pi)$ factors. The first order is 
\begin{align}
	H_I^{(1)}=\frac{1}{T}\int_0^TH_I(s)ds&=\left(2\pi\right)\sum_k^N\left[\left(A_{\rm CS}^k+\Delta^k\right)I_S^k+\Omega^kI_A^k +\Omega^k\frac{\sin{\left(4\pi \omega^kT\right)}I_A-\left(\cos{\left(4\pi \omega^kT\right)}-1\right)I_M}{4\pi \omega^k}\right].\nonumber
\end{align}
If we assume $T=p/\omega^k$, being $p$ integer,
\begin{align}
	H_I^{(1)}=\left(2\pi\right)\sum_k^N\left[\left(A_{\rm CS}^k+\Delta^k\right)I_S^k+\Omega^kI_A^k\right].
\end{align}
The second order contribution is given by 
\begin{align}
	H_I^{(2)}=\frac{1}{2iT}\int_0^Tdt\int_0^t\left[H_I(t),H_I(s)\right]ds,\nonumber
\end{align}
and the commutator gives the following non-zero terms
\begin{align}
(1)\rightarrow&-\frac{\left(2\pi\right)^2}{2}\left(A_{\rm CS}+\Delta^k\right)\left(\Omega^k\right)\left[\cos{\left(2\pi\;2\omega^ks\right)}\left(iI_M\right)+\sin{\left(2\pi\;2\omega^ks\right)}\left(iI_A\right)\right]+\frac{\left(2\pi\right)^2}{2}\left(\Omega^k\right)^2\sin{\left(2\pi\;2\omega^ks\right)}\left(-iI_S\right)\nonumber\\[8pt]
(2)\rightarrow&-\frac{\left(2\pi\right)^2}{2}\left(A_{\rm CS}+\Delta^k\right)\left(\Omega^k\right)\left[-\cos{\left(2\pi\;2\omega^kt\right)}\left(iI_M\right)-\sin{\left(2\pi\;2\omega^kt\right)}\left(iI_A\right)\right]+\frac{\left(2\pi\right)^2}{2}\left(\Omega^k\right)^2\sin{\left(2\pi\;2\omega^kt\right)}\left(iI_S\right)\nonumber\\[8pt]
(3)\rightarrow&\frac{1}{2}\left(\Omega^k\right)^2\sin{\left(2\pi\;2\omega^k\left(t-s\right)\right)}\left(i I_S\right).\nonumber
\end{align}
And now let us compute the result of the integral for each of these terms
\begin{align}
(1)\rightarrow&\frac{1}{2iT}\int_0^Tdt\int_0^s\left\{\frac{-\left(2\pi\right)^2}{2}\left(A_{\rm CS}+\Delta^k\right)\left(\Omega^k\right)\left[\cos{\left(2\pi\;2\omega^ks\right)}\left(iI_M\right)+\sin{\left(2\pi\;2\omega^ks\right)}\left(iI_A\right)\right]+\frac{\left(2\pi\right)^2}{2}\left(\Omega^k\right)^2\sin{\left(2\pi\;2\omega^ks\right)}\left(-iI_S\right)\right\}ds\nonumber\\[6pt]
&=-\frac{1}{4}\left(A_{\rm CS}+\Delta^k\right)\left(\Omega^k\right)\left[\frac{\sin^2{\left(2\pi \omega^k T\right)}}{2\pi\left(\omega^k\right)^2T}I_M-\frac{\sin{\left(2\pi\;2\omega^kT\right)}-4\pi \omega^kT}{4\left(\omega^k\right)^2T}I_A\right]+\frac{1}{4}\left(\Omega^k\right)^2\frac{\sin{\left(2\pi\;2\omega^kT\right)}-4\pi \omega^kT}{4\left(\omega^k\right)^2T}I_S\nonumber\\[8pt]
(2)\rightarrow&\frac{1}{2iT}\int_0^Tdt\int_0^s\left\{-\frac{\left(2\pi\right)^2}{2}\left(A_{\rm CS}+\Delta^k\right)\left(\Omega^k\right)\left[\cos{\left(2\pi\;2\omega^kt\right)}\left(-iI_M\right)-\sin{\left(2\pi\;2\omega^kt\right)}\left(iI_A\right)\right]+\frac{\left(2\pi\right)^2}{2}\left(\Omega^k \right)^2\sin{\left(2\pi\;2\omega^kt\right)}\left(iI_S\right)\right\}ds\nonumber\\[6pt]
&=-\frac{1}{4}\left(A_{\rm CS}+\Delta^k\right)\left(\Omega^k\right)\left[\frac{-\cos{\left(4\pi \omega^kT\right)}+1-4\pi \omega^kT\sin{\left(4\pi \omega^kT\right)}}{4\left(\omega^k\right)T}I_M+\frac{4\pi \omega^kT\cos{\left(4\pi \omega^kT\right)}-\sin{\left(4\pi \omega^kT\right)}}{4\left(\omega^k\right)^2T}I_A\right]\nonumber\\[6pt]
&+\frac{1}{4}\left(\Omega^k\right)^2\frac{\sin{\left(4\pi \omega^kT\right)}-4\pi \omega^k T\cos{\left(4\pi \omega^k T\right)}}{4 \left(\omega^k\right)^2 T}I_S\nonumber\\[8pt]
(3)\rightarrow&\frac{1}{2iT}\int_0^Tdt\int_0^s\left(2\pi\right)^2\frac{1}{2}\left(\Omega^k\right)^2\sin{\left(2\pi\;2\omega^k\left(t-s\right)\right)}\left(i I_S\right)=\left(\Omega^k\right)^2\frac{4\pi \omega^k T-\sin{\left(4\pi \omega^k T\right)}}{8\left(\omega^k\right)^2T}I_S\nonumber
\end{align}
By picking the periods such that $T=p/\omega^k$, then the total contribution of this second order Hamiltonian will be (we divide it by $2\pi$ so that everything is written back in Hz)
\begin{align}
H_I^{(2)}=-\frac{\left(A_{\rm CS}+\Delta^k\right)\left(\Omega^k\right) }{4 \omega^k}I_A.\nonumber
\end{align}
Assuming that $A_{\rm CS}+\Delta^k\approx \Delta_k=\frac{\Omega^k}{\sqrt{2}}$, one gets
\begin{align}
H_I^{(2)}=-\frac{\left(\Omega^k\right)^2}{4\sqrt{2}\omega^k}\;I_A.\nonumber
\end{align}
Then, the total Hamiltonian with the first and second order contributions reads
\begin{align}
H\approx H_I^{(1)}+H_I^{(2)}=\sum_k^N&\Bigg\{\left(A_{\rm CS}^k+\Delta^k\right)I_S^k+\Omega^kI_A^k+ \frac{\left(\Omega^k\right)^2}{4\sqrt{2}\omega^k}\;I_A^k\Bigg\}.\nonumber
\end{align}
Then, in the IP w.r.t. the effective driving, we obtain the modified dynamics
\begin{align}
H_{II}\approx\sum_{k}^N\frac{1}{\sqrt{3}}\left[\delta_{\rm iso}^{(k)}+\frac{\left(\Omega^k\right)^2}{4\omega^k}\right]I_{\hat{n}}^k,\label{eq:appB:Havg}
\end{align}
where we took the average over a full magnetic field cycle, $\overline{A_{\rm CS}^k}=\delta_{\rm iso}^{(k)}$.
Note that the Bloch-Siegert effect induces a shift to the isotropic chemical shift. Such shift depends on the amplitude of the applied RF driving and the strength of its frequency $\omega^k$. 

\section{Initial state of the sample and emitted signal.}\label{app:signal}

\textit{\textbf{Sample initial state.}}

 Let us discuss the initial state of the sample. In practice, each spin within the detection volume will be in an almost random pure state, so that the average state of the detection volume is
\begin{align}
    \rho_{\rm sample}&=\frac{1}{N}\sum_k^N\left[\cos^2\left(\frac{\theta_k}{2}\right)|0\rangle\langle0|+\sin^2\left(\frac{\theta_k}{2}\right)|1\rangle\langle1|+\cos\left(\frac{\theta_k}{2}\right)\sin\left(\frac{\theta_k}{2}\right)e^{i\varphi_k}|0\rangle\langle1|+\cos\left(\frac{\theta_k}{2}\right)\sin\left(\frac{\theta_k}{2}\right)e^{-i\varphi_k}|1\rangle\langle0|\right]\nonumber\\[8pt]
    &=\frac{1}{N}\sum_k^N\left[\frac{\mathds{1}}{2}+\cos{\theta_k}\ I_{\hat{n}}+\sin{\theta_k}\cos{\varphi_k}\;I_{\hat{u}}+\sin{\theta_k}\sin{\varphi_k}\;I_{\hat{v}}\right],\label{eq:appB:s_state0}
\end{align}
where we retain only the linear combination of Pauli matrices, since higher-order terms do not usually contribute to the observed signal. Note that we use the arbitrary directions \(\{\hat{u},\hat{v},\hat{n}\}\) defined by the magnetic field orientation (see the previous section).

In the macroscopic case limit ($N\rightarrow\infty$), the sample mean state is the thermal state
\begin{align}
    \rho_{\rm th}=e^{\beta\gamma B_0}|0\rangle\langle 0|+e^{-\beta\gamma B_0}|1\rangle\langle 1|,\nonumber
\end{align}
with $\beta=\frac{\hbar}{K T}$, being $K$ the Boltzmann constant and $T$ temperature.
Let us find the probability distribution for \(\theta_k\) that gives rise to the thermal state in the macroscopic limit. Following the standard thermodynamic approach, we assume that the spin system is in contact with a heat reservoir. Fixing the system in a given microstate \(k\), the total energy is \(E_{\rm tot}=E_{\rm res}+E_k\), so that \(E_{\rm res}=E_{\rm tot}-E_k\). Note that degeneracies may occur, meaning that different microstates can correspond to the same energy.

Once the microstate \(k\) is fixed, the probability of finding the system in \(k\) is proportional to the number of ways of arranging the combined system-reservoir ensemble. Since \(E_{\rm tot}\) is fixed, this is equivalent to counting the number of ways in which the reservoir can be arranged:
\begin{align}
    P_k=P_{\rm res}(E_{\rm res})=C\times\Omega_{\rm res}(E_{\rm res})=C\times\Omega_{\rm res}(E_{\rm tot}-E_k),\nonumber
\end{align}
where $\Omega_{\rm res}$ is the number of microstates of the reservoir given its energy $E_{\rm res}$. Assuming that $E_k\ll E_{\rm res}\approx E_{\rm tot}$, we can expand $\ln{\left(\Omega_{\rm res}(E_{\rm tot}-E_k)\right)}$ around $E_k=0$
\begin{align}
    \ln{\left(\Omega_{\rm res}(E_{\rm tot}-E_k)\right)}=\ln{\left(\Omega_{\rm res}(E_{\rm tot})\right)}-E_k\frac{\partial\ln{\left(\Omega_{\rm res}(E)\right)}}{\partial E}+...\nonumber
\end{align}
Where $\frac{\partial\ln{\left(\Omega_{\rm res}(E)\right)}}{\partial E}=\beta$ if the number of particles of the reservoir is large, $N_{\rm res}\rightarrow\infty$. Thus,
\begin{align}
    \Omega_{\rm res}(E_{\rm tot}-E_k)=\Omega_{\rm res}(E_{\rm tot})e^{-\beta E_k}\longrightarrow P_k=\frac{1}{Z}e^{-\beta E_k},\nonumber
\end{align}
where $Z$ is for now a normalization constant. We found the Boltzmann distribution without imposing any constraint to the system's size (only assumed a large reservoir). To impose normalization and find $Z$, let us specify a little bit more the system. We are treating a spin $1/2$ system, which in general is aligned in a direction that forms an angle $\theta_k$ ($0<\theta_k<\pi$) with respect to the external magnetic field $B_0$. Thus, $$E_k=\gamma B_0\left(\cos^2{\theta_k/2}-\sin^2{\theta_k/2}\right)=\gamma B_0\cos{\theta_k}.$$ Now, let us impose normalization, $\displaystyle\sum_kP_k=1$. The sum over all possible microstates $k$ can be translated into an integral over all possible orientations, $\displaystyle\sum_k\longrightarrow\int_0^\pi \sin{\theta}\ d\theta$. This leads to 
\begin{align}
    Z=\frac{2\sinh{\left(\beta\gamma B_0\right)}}{\beta\gamma B_0},\nonumber
\end{align}
and thus to 
\begin{align}
    P_k\equiv P(\theta)=\frac{\beta\gamma B_0}{2\sinh{\left(\beta\gamma B_0\right)}}e^{-\beta\gamma B_0\cos{\theta}}.\label{eq:appB:distr}
\end{align}
The cumulative probability of a spin lying in the $[0,\theta]$ range is 
\begin{align}
    P_c(\theta)=\int_0^\theta P(\theta')\sin{\theta'}d\theta'=\frac{e^{-\beta\gamma B_0\cos{\theta}}-e^{-\beta\gamma B_0}}{2\sinh{\left(\beta\gamma B_0\right)}}.\nonumber
\end{align}
Thus we have the probability of a given spin to fall into the $[0,\theta]$ range. With this distribution, Eq.~\eqref{eq:appB:s_state0} tends to the Boltzmann distribution for a large number of spins.
This is a more fundamental way of understanding the origin of statistical polarisation, which gives rise to a the random initial phase detected at nanoscale scenarios.

\vspace{1em}
\textit{\textbf{Coupling to the sensors and effective emitted signal.}}

Recalling from Appendix\ref{app:dyn_ensemble} and Eq.~\eqref{eq:traj} in the main text, the effective sample Hamiltonian is
\begin{align}
    H_{\rm tot}=\frac{1}{\:\sqrt{3}}\sum_{k=1}^N\delta_{\rm iso}^k\:I_{\hat{n}}^k.\nonumber
\end{align}
The initial state of the sample is given by Eq.~\eqref{eq:appB:s_state0}. Since our goal is to detect statistical polarisation, there is no need of the usual initialisation $\left(\pi/2\right)$-pulse. 
After an evolution time $\tau$ under $H_{\rm tot}$, the state is
\begin{align}
    \rho_{\rm sample}(\tau)=\sum_{k=1}^N\left[\frac{\mathds{1}_k}{2}+\cos{\left(\theta_k\right)}I_{\hat{n}}^k+\sin{\left(\theta_k\right)}\sin{\left(\frac{\delta_{\rm iso}^k}{\sqrt{3}}\tau+\varphi_k\right)}I_{\hat{u}}^k+\sin{\left(\theta_k\right)}\cos{\left(\frac{\delta_{\rm iso}^k}{\sqrt{3}}\tau+\varphi_k\right)}I_{\hat{v}}^k\right].\label{eq:appC:s_state3}
\end{align}

The oscillations of frequency $\frac{\delta_{\rm iso}^k}{\sqrt{3}}$ will be detected by the corresponding NV through the dipole-dipole interaction
\begin{align}
    H_{\rm dd}=\frac{\mu_0 \gamma_e h}{4\pi}\sum_k^N\frac{\gamma_k}{\lvert\vec{r}^{\ k}\rvert^3}  \left[\vec{S}\cdot\vec{I}^{\ k}-3\left(\vec{S}\cdot\hat{l}^{k}\right)\left(\vec{I}^{\ k}\cdot\hat{l}^{k}\right)\right],\nonumber
\end{align}
where $\vec{r}$ is the vector from the NV center to the nucleus in the sample, $\hat{l}^{k}=\frac{\vec{r}^{\ k}}{\lvert\vec{r}^{\ k}\rvert}$ is the corresponding unit vector, $\gamma_e$ and $\gamma_k$ are the electron and nuclear gyromagnetic ratios, and $\vec{S}$ and $\vec{I}^{\ k}$ are the electron (NV) and nuclear spin operators, respectively. 

We impose that the interrogation windows happen only once every full magnetic field rotation, $\nu \tau=p\left(2\pi\right)$ (for any integer $p$), so that the dynamics are exactly given by $H_{\rm tot}$ above. In addition, since during the interrogation windows $\nu\tau=p\left(2\pi\right)$, the axes defined by Eqs.~\eqref{eq:app:osc_axes} are
\begin{align}
I_{M(t)}^k&=I_{\hat{u}}^k=I_x^k\nonumber\\[8pt]
I_{A(t)}^k&=\frac{1}{\sqrt{3}}\:I_{\hat{v}}^k+\sqrt{\frac{2}{3}}\:I_{\hat{n}}^k=I_y^k,\nonumber\\[8pt]
I_{S(t)}^k&=-\sqrt{\frac{2}{3}}\:I_{\hat{v}}^k+\frac{1}{\sqrt{3}}\:I_{\hat{n}}^k=I_z^k.
\end{align}

In the interaction picture (I.P.) with respect to the free Hamiltonians of the NV and the nuclei,  $H_0=\omega_{\rm NV} S_S+\omega\sum_{k=1}^N I_S^k$, the interaction Hamiltonian in the secular approximation reduces to
\begin{align}
    H_{\rm dd}=\frac{\mu_0 \gamma_e h}{4\pi}\sum_k^N\frac{\gamma_k}{\lvert\vec{r}^{\;k}\rvert^3} S_{S} \left[3l_s^{k}l_m^{k},3l_s^{k}l_a^{k},3\left(l_s^{k}\right)^2-1\right]\cdot\left[I_M^{k}\cos{\left(\omega t\right)}-I_A^{k}\sin{\left(\omega t\right)},I_A^{k}\cos{\left(\omega t\right)}+I_M^{k}\sin{\left(\omega t\right)},I_S^{k}\right],\nonumber
\end{align}
where \(\{l_m^k,l_a^k,I_s^k\}\) are the geometric unit vectors pointing from the NV center to each spin. $\omega$ is the Larmor frequency of the nucleus. In this same IP, the driving is, recall from Appendix\ref{app:dyn_ensemble},
\begin{align}
    H_c=\sum_k^N\sqrt{\frac{2}{3}}\Omega\left[I_{S}^k+\sqrt{2}\ I_{A}^k\right]=\sum_{k=1}^N\tilde{\Omega}\ I_{\hat{n}},\nonumber
\end{align}
with $\tilde{\Omega}=\sqrt{2}\Omega$.

Now, we would like to see the effect of the driving on the dipolar interaction. Notice that the contributions with $I_M^k$ and $I_A^k$ can be safely neglected in the secular approximation. Let us see how the remaining contribution $I_z^k$ is affected by $H_c$. Since we impose $\nu\tau=p\left(2\pi\right)$, $I_S^k=-\sqrt{\frac{2}{3}}I_v^k+\frac{1}{\sqrt{3}}\:I_n^k=I_z^k$, the dipolar term in the IP with respect to $H_c$ is
\begin{align}
    H_{\rm dd}=-\frac{\mu_0 \gamma_e h}{4\pi}\sum_k^N\frac{\gamma_k}{\lvert\vec{r}^{\;k}\rvert^3}  \left[3\left(l_z^{k}\right)^2-1\right]\sqrt{\frac{2}{3}}S_z\left[\cos{\left(\tilde{\Omega}^k t\right)}\;I_v^k+\sin{\left(\tilde{\Omega}^k t\right)}I_u^k\right],\nonumber
\end{align}
where we neglected the $I_n^k$ contribution since it will be averaged out one the pulsed scheme is applied.

Now, we assume that the spin operators of the nuclei can be safely approximated by their expectation values, valid when considering that the sample is composed of $N\sim 10^4$ spins. This allows us to write the dipolar Hamiltonian in the following compact form
\begin{align}
    H_{\rm dd}(t)=\gamma_eB_N(t)S_z,\label{eq:appC:eff_sig}
\end{align}
with
\begin{align}
    B_N(t)&=-\frac{\mu_0h}{4\pi}\sum_k^N\gamma_k\sqrt{\frac{2}{3}}\left[\frac{3\left(l_z^k\right)^2-1}{\lvert\vec{r}^{\ k}\rvert^3}\right]\Biggl[\cos{\left(\tilde{\Omega}^k t\right)}\langle I_u^k\rangle+\sin{\left(\tilde{\Omega}^k t\right)}\langle I_v^k\rangle\Biggr]=\frac{\mu_0h}{4\pi}\sum_k^N\gamma_kC_k\left[\cos{\left(\tilde{\Omega}^k t\right)}\langle I_v^k\rangle+\sin{\left(\tilde{\Omega}^k t\right)}\langle I_u^k\rangle\right].\nonumber
\end{align}
From Eq.~\eqref{eq:appC:s_state3}, we have
\begin{align}
    &\langle I_u^k\rangle=\frac{1}{2}\sin{\left(\theta_k\right)}\sin{\left(\frac{\delta_{\rm iso}^k}{\sqrt{3}}\tau+\varphi_k\right)},\hspace{1.5em}\langle I_v^k\rangle=\frac{1}{2}\sin{\left(\theta_k\right)}\cos{\left(\frac{\delta_{\rm iso}^k}{\sqrt{3}}\tau-\varphi_k\right)}.\nonumber
\end{align}
Assuming all the spins are of the same species ($\gamma_k\rightarrow\gamma$ and $\Omega^k\rightarrow\Omega$) this leads to
\begin{align}
B_N(t)&=\frac{\mu_0\gamma h}{4\pi}\frac{1}{2}\sum_{k=1}^N\left\{\sin{\left(\theta_k\right)}C_k\left[\cos{\left(\tilde{\Omega} t\right)}\cos{\left(\frac{\delta_{\rm iso}^k}{\sqrt{3}}\tau+\varphi_k\right)}+\sin{\left(\tilde{\Omega} t\right)}\sin{\left(\frac{\delta_{\rm iso}^k}{\sqrt{3}}\tau+\varphi_k\right)}\right]\right\}\nonumber\\[8pt]
&=\frac{\mu_0\gamma h}{4\pi}\frac{1}{2}\sum_{k=1}^N\left\{\sin\left(\theta_k\right)C_k\cos{\left(\tilde{\Omega} t-\frac{\delta_{\rm iso}}{\sqrt{3}}\tau-\varphi_k\right)}\right\}\label{eq:appC:eff_Bfield}
\end{align}
Note that this term increases its magnitude as $\sqrt{N}$. This can be seen from the standard deviation
\begin{align}
    \delta B_N(t)&=\sqrt{{\rm var}\left[B_N(t)\right]}=\frac{\mu_0 \gamma h}{8\pi}\sqrt{N}\left\{\int\left[\frac{3\left(l_z\right)^2-1}{\lvert\vec{r}\rvert^3}\right]^2dV\times\int_0^\pi\sin^3{\left(\theta\right)P(\theta)}d\theta\times\frac{1}{2\pi}\int_0^{2\pi}\cos^2{\left(\tilde{\Omega}t-\frac{\delta_{\rm iso}}{\sqrt{3}}\tau-\varphi\right)}d\varphi\right\}^{1/2}\nonumber\\[8pt] 
    &=\frac{\mu_0 \gamma h}{8\pi}\sqrt{N}\sqrt{F_z^{(2)}\times\frac{2}{3}}\nonumber,
\end{align}

where we defined $\displaystyle F_q^{(2)}=\int \left[\frac{3\left(n_z\right)^2-1}{\lvert \vec{\;r}\rvert^3}\right]^2dV$. However, the mean of such term vanishes when performing the average over different detection volumes (different NVs) or experimental shots, 
\begin{align}
    \overline{B_N^q(t)}\approx&\frac{\mu_0\gamma h}{4\pi}\frac{1}{2}\sum_{k=1}^N\left\{C_k^q\times\int_0^\pi\sin^2{\left(\theta_k\right)P(\theta_k)}d\theta_k\right.\nonumber\\[8pt]
    &\left.\times\frac{1}{2\pi}\int_0^{2\pi}\left[\cos{\left(\tilde{\Omega} t\right)}\cos{\left(\frac{\delta_{\rm iso}^k}{\sqrt{3}}\tau+\varphi_k\right)}+\sin{\left(\tilde{\Omega} t\right)}\sin{\left(\frac{\delta_{\rm iso}^k}{\sqrt{3}}\tau+\varphi_k\right)}\right]d\varphi_k\right\}=0.\nonumber
\end{align}
For a single shot, the signal has a mean amplitude given by the standard deviation $\delta B_N^q(t)$, and has a general form of 
\begin{align}
B_N^{\rm single}(t)=\frac{\mu_0\gamma h}{8\pi}\sqrt{N}\sqrt{\frac{2}{3}}\sqrt{F_z^{(2)}}\left[\cos{\left(\tilde{\Omega} t\right)}\cos{\left(\frac{\delta_{\rm iso}}{\sqrt{3}}\tau+\varphi\right)}+\sin{\left(\tilde{\Omega} t\right)}\sin{\left(\frac{\delta_{\rm iso}}{\sqrt{3}}\tau+\varphi\right)}\right],
\end{align}
where $\varphi$ might change between sensing volumes of experimental shots.
\section{NV detection}\label{app:sensor}
First, the electron sensor and the memory qubit (neighbouring niteogen) are initialized. Here, we will use $\vec{S}$ and $\vec{P}$ to denote the spin operators of the electron and the nitrogen spin. The nitrogen is assumed to be $^{15}{\rm N}$, so that $P=1/2$, although the protocol would work similarly in the case of workng with $^{14}{\rm N}$ ($P=1$).  We will assume no coupling between the different sensors in the ensemble, so we treat them as independent. The total Hilbert space is then $\mathcal{H}_e\otimes\mathcal{H}_n$. For the moment, we are going to consider the case of high magnetic field, so that the nitrogen spin is not affected by the green laser, and that the probing of the sample steps are only applied once one every period $T=1/\nu$ of the magnetic field, so that the magnetic field direction is always aligned with the NV axis and, thus, we avoid mixing of the levels during initialisation. This implies that $q\rightarrow z$ in Eq.~\eqref{eq:appC:eff_sig}.

The initial state after initialisation is

\begin{align}
    \rho_0(0)=\left[\frac{\mathds{1}}{2}+S_z\right]\left[\frac{\mathds{1}}{2}+P_z\right].\nonumber
\end{align}
A microwave (MW)$(\pi/2)_x$-pulse is applied, leading to 
\begin{align}
    \rho(0)=\left[\frac{\mathds{1}}{2}-S_y\right]\left[\frac{\mathds{1}}{2}+P_z\right].\nonumber
\end{align}
 Now, the sample is probed by the electrons. Remember that we assume that sample state in each detection volume is governed by statistical polarisation. That is, Eq.~\eqref{eq:appC:eff_Bfield} is probed through the effective interaction Eq.~\eqref{eq:appC:eff_sig}. With $\tau=0$ and assuming $\nu\ll\Omega_{\rm LG}$, we consider $\nu t\approx 2\pi k$ for all the probing time $t_{\rm prob}$, and using a XY4 MW pulsed scheme with pulse spacing $T_{\rm pulse}=\frac{1}{2\Omega_{\rm LG}}$, one obtains
 \begin{align}
     \rho_1(0)=\left[\frac{\mathds{1}}{2}-S_y\cos{\left(\frac{2\gamma_e t_{\rm prob}}{\pi}g\sum_{k=1}^N\left[C_k^z\sin{\left(\theta_k\right)}\sin{\left(-\varphi_k\right)}\right]\right)}+S_x\sin{\left(\frac{2\gamma_e t_{\rm prob}}{\pi}g\sum_{k=1}^N\left[C_k^z\sin{\left(\theta_k\right)}\sin{\left(-\varphi_k\right)}\right]\right)}\right]\Bigg[\frac{\mathds{1}}{2}+P_z\Bigg],\nonumber
 \end{align}
 where $g=\frac{\mu_0\gamma h}{8\pi}$ and $t_{\rm prob}=\frac{2}{\Omega_{\rm LG}}$. The information is now transferred to the populations of the NV with a $\left(-\pi/2\right)_y$ pulse and then, via a $\rm{C_eNOT_n}$ gate, to the state of the memory qubit, leading to 
\begin{align}
    \rho_2(0)=\frac{\mathds{1}}{4}+\frac{\left(S_z+P_z\right)}{2}\sin{\left(\frac{2\gamma_e t_{\rm prob}}{\pi}g\sum_{k=1}^N\left[C_k^z\sin{\left(\theta_k\right)}\sin{\left(\varphi_k\right)}\right]\right)}-\left(S_yP_x-S_xP_y\right)\cos{\left(\frac{2\gamma_e t_{\rm prob}}{\pi}g\sum_{k=1}^N\left[C_k^z\sin{\left(\theta_k\right)}\sin{\left(\varphi_k\right)}\right]\right)}+S_zP_z.\nonumber
\end{align}
Note that if we trace out the NV state, the remaining state of the memory qubit is
\begin{align}
    \rho_2^{\rm memory}(0)=\frac{\mathds{1}}{2}+P_z\sin{\left(\frac{2\gamma_e t_{\rm prob}}{\pi}g\sum_{k=1}^N\left[C_k^z\sin{\left(\theta_k\right)}\sin{\left(\varphi_k\right)}\right]\right)}.\nonumber
\end{align}
Thus, the information about the initial state is correctly stored in the populations of the memory qubit.

Now the squared block (to be repeated $M$ times) depicted in Fig.~\ref{fig:sequence} (Top left) starts. First, the sample is evolved for a time $\tau$ which, recall, we set for the moment equal to the period of the magnetic field, $\tau=1/\nu$. After such evolution, the NV is again initialized and sent to the xy plane by means of a $\left(\pi/2\right)_x$ pulse. Now it probes the sample again, leading to
\begin{align}
    \rho_1(\tau)=&\left[\frac{\mathds{1}}{2}-S_y\cos{\left(\frac{2\gamma_e t_{\rm prob}}{\pi}g\sum_{k=1}^N\left[C_k^z\sin{\left(\theta_k\right)}\sin{\left(\frac{\delta_{\rm iso}^k}{\sqrt{3}}\tau+\varphi_k\right)}\right]\right)}+S_x\sin{\left(\frac{2\gamma_e t_{\rm prob}}{\pi}g\sum_{k=1}^N\left[C_k^z\sin{\left(\theta_k\right)}\sin{\left(\frac{\delta_{\rm iso}^k}{\sqrt{3}}\tau+\varphi_k\right)}\right]\right)}\right]\nonumber\\[8pt]
    &\times\Bigg[\frac{\mathds{1}}{2}+P_z\sin{\left(\frac{2\gamma_e t_{\rm prob}}{\pi}g\sum_{k=1}^N\left[C_k^z\sin{\left(\theta_k\right)}\sin{\left(\varphi_k\right)}\right]\right)}\Bigg].\nonumber
\end{align}
Now, we apply a $(-\pi/2)_x$ pulse over the NV and then recover the information stored in the memory qubit by means of a ${\rm C_nNOT_e}$ gate, leading to the state
\begin{align}
    \rho_2(\tau)&=\frac{\mathds{1}}{4}+S_zP_z\sin{\left(\frac{2\gamma_e t_{\rm prob}}{\pi}g\sum_{k=1}^N\left[C_k^z\sin{\left(\theta_k\right)}\sin{\left(\frac{\delta_{\rm iso}^k}{\sqrt{3}}\tau+\varphi_k\right)}\right]\right)}-S_yP_z\cos{\left(\frac{2\gamma_e t_{\rm prob}}{\pi}g\sum_{k=1}^N\left[C_k^z\sin{\left(\theta_k\right)}\sin{\left(\frac{\delta_{\rm iso}^k}{\sqrt{3}}\tau+\varphi_k\right)}\right]\right)}\nonumber\\[8pt]
    &-\frac{S_y}{2}\cos{\left(\frac{2\gamma_e t_{\rm prob}}{\pi}g\sum_{k=1}^N\left[C_k^z\sin{\left(\theta_k\right)}\sin{\left(\varphi_k\right)}\right]\right)}\cos{\left(\frac{2\gamma_e t_{\rm prob}}{\pi}g\sum_{k=1}^N\left[C_k^z\sin{\left(\theta_k\right)}\sin{\left(\varphi_k\right)}\right]\right)}\nonumber\\[8pt]
    &+\frac{S_z}{2}\sin{\left(\frac{2\gamma_e t_{\rm prob}}{\pi}g\sum_{k=1}^N\left[C_k^z\sin{\left(\theta_k\right)}\sin{\left(\varphi_k\right)}\right]\right)}\sin{\left(\frac{2\gamma_e t_{\rm prob}}{\pi}g\sum_{k=1}^N\left[C_k^z\sin{\left(\theta_k\right)}\sin{\left(\varphi_k\right)}\right]\right)}\nonumber\\[8pt]
    &+\frac{P_z}{2}\sin{\left(\frac{2\gamma_e t_{\rm prob}}{\pi}g\sum_{k=1}^N\left[C_k^z\sin{\left(\theta_k\right)}\sin{\left(\varphi_k\right)}\right]\right)}.\label{eq:appC:corr_state}
\end{align}
Note that 
\begin{align}
    \langle S_z\rangle&=\frac{1}{2}\sin{\left(\frac{2\gamma_e t_{\rm prob}}{\pi}g\sum_{k=1}^N\left[C_k^z\sin{\left(\theta_k\right)}\sin{\left(\frac{\delta_{\rm iso}^k}{\sqrt{3}}\tau+\varphi_k\right)}\right]\right)}\sin{\left(\frac{2\gamma_e t_{\rm prob}}{\pi}g\sum_{k=1}^N\left[C_k^z\sin{\left(\theta_k\right)}\sin{\left(\varphi_k\right)}\right]\right)}\nonumber\\[8pt]
    &\approx\frac{1}{2}\left(\frac{2\gamma_e t_{\rm prob}}{\pi}g\right)^2\sum_{k=1}^N\left[C_k^z\sin{\left(\theta_k\right)}\sin{\left(\frac{\delta_{\rm iso}^k}{\sqrt{3}}\tau+\varphi_k\right)}\right]\sum_{l=1}^N\left[C_l^z\sin{\left(\theta_l\right)}\sin{\left(\varphi_l\right)}\right],\nonumber
\end{align}
where we assumed $\frac{2\gamma_e t_{\rm prob}}{\pi}g\ll 1$. Let us work a little bit in this expression. The double sum is divided into diagonal elements ($k=l$) and non-diagonal ($k\neq l$)
\begin{align}
    \frac{1}{2}\left(\frac{2\gamma_e t_{\rm prob}}{\pi}g\right)^2\left\{\sum_{k=1}^N\left[\left(C_k^z\right)^2\sin^2{\left(\theta_k\right)}\sin{\left(\frac{\delta_{\rm iso}^k}{\sqrt{3}}\tau+\varphi_k\right)}\sin{\left(\varphi_k\right)}\right]+\sum_{k\neq l}^N\left[C_k^zC_l^z\sin{\left(\theta_k\right)}\sin{\left(\theta_l\right)}\sin{\left(\frac{\delta_{\rm iso}^k}{\sqrt{3}}\tau+\varphi_k\right)}\sin{\left(\varphi_l\right)}\right]\right\}\nonumber
\end{align}
The non-diagonal term can be disregarded, since when performing the average will vanish. By taking the ensemble average of the first term, recalling Eq.~\eqref{eq:appB:distr}, one gets

\begin{align}
    &\overline{\langle S_z}\rangle\approx \frac{1}{2}\left(\frac{2\gamma_e t_{\rm prob}}{\pi}g\right)^2\sum_{k=1}^N\left[\cos{\left(\frac{\delta_{\rm iso}^k}{\sqrt{3}}\tau\right)}\int \left(\frac{3\left(n_z^k\right)^2-1}{\lvert \vec{\;r}^{\;k}\rvert^3}\right)^2dV\times\frac{1}{2\pi}\int_0^{\pi}\sin^3{\left(\theta_k\right)}P(\theta_k)d\theta_k\int_0^{2\pi}\left[\sin{\left(\frac{\delta_{\rm iso}^k}{\sqrt{3}}\tau+\varphi_k\right)}\sin{\left(\varphi_k\right)}\right] d\varphi_k \right]\nonumber\\[8pt]
    &=\frac{1}{2}\left(\frac{2\gamma_e t_{\rm prob}}{\pi}g\right)^2\frac{F_z^{(2)}}{3}\sum_{k=1}^N\cos{\left(\frac{\delta_{\rm iso}^k}{\sqrt{3}}\tau\right)}\nonumber
\end{align}
For the general case where the $N$ nuclei are partitioned into $l$ non-equivalent groups, with group $i$ containing $N^{(i)}$ spins (such that $\displaystyle\sum_{i=1}^l N^{(i)} = N$), the signal is
\begin{align}
    \langle \overline{S_z}\rangle=\frac{1}{2}\left(\frac{2\gamma_e t_{\rm prob}}{\pi}g\right)^2\frac{F_z^{(2)}}{3}\sum_{i=1}^lN^{(i)}\cos{\left(\frac{\delta_{\rm iso}^i}{\sqrt{3}}\tau\right)}.\nonumber
\end{align} 

\newpage
\section{Simple driving and misalignment errors}\label{app:misa}
Let us consider a simpler version of the driving. Following the same steps of Appendix~\ref{app:dyn_ensemble}, it is easy to show that a simplified version of the driving , such as
\begin{align}
        H_c(t)&=2\Omega\cos\left(\omega t+\alpha\right)\sum_{k=1}^NI_{\hat{n}}^k,\label{app:eq:simdriv}
\end{align}
leads to the same sample dynamics.

To consider misalignment errors, we follow the same steps as in the main text. We consider a modified version of Eq.~\eqref{app:eq:simdriv}
\begin{align}
        H_c(t)&=2\Omega\cos\left(\omega t+\alpha\right)\sum_{k=1}^N\left[aI_{\hat{n}}^k+bI_{\hat{u}}^k+cI_{\hat{v}}^k\right].\label{eq:app:errdriving}
\end{align}
The imperfections are modeled by starting from the ideal direction vector
\(
\mathbf{v}_0=\hat{n},
\)
and introducing a random unit vector \(\mathbf{u}\) orthogonal to \(\mathbf{v}_0\). The misaligned direction is then written as
\(
\mathbf{v}=\cos(\phi_{\rm error})\,\mathbf{v}_0+\sin(\phi_{\rm error})\,\mathbf{u},
\)
so that the total angle between \(\mathbf{v}\) and the ideal case is exactly \(\phi_{\rm error}\). In this way, the coefficients in the driving Hamiltonian are given by
\begin{align}
	a=v_1,\hspace{5mm}
	b=v_2,\hspace{5mm}
	c=v_3.\nonumber
\end{align}

\begin{figure*}[]
\includegraphics[width=0.6 \linewidth]{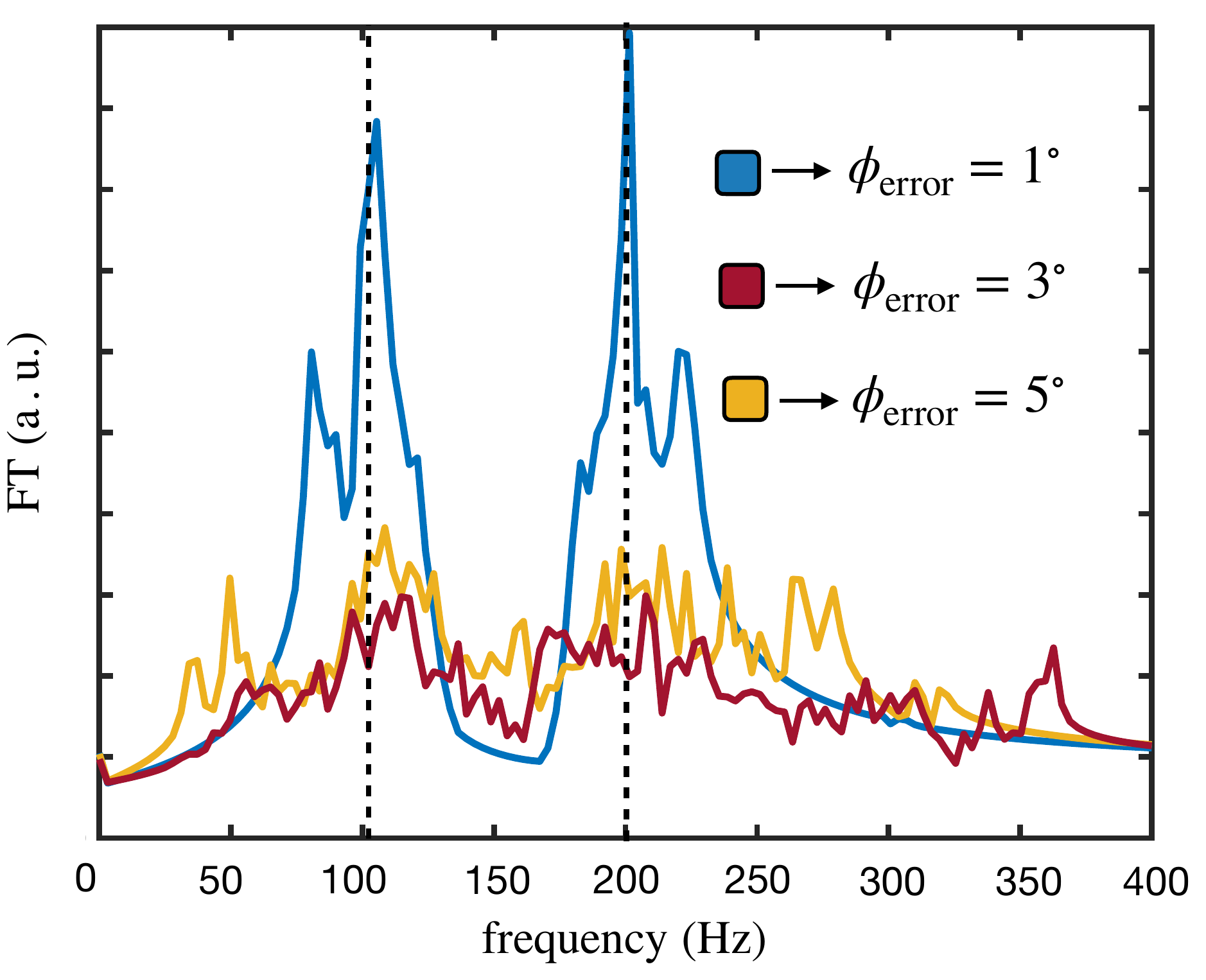}
\caption{Simulated spectrum of the sample with two magnetically nonequivalent protons. The colored solid lines correspond to the simulated spectrum obtained with the version of the driving given by Eq.~\eqref{eq:app:errdriving} for different misalignment errors. The dashed vertical lines indicate the peak positions predicted by the theory.  All the simulations also include stochastic amplitude errors of 0.25\% with correlation time of 1 ms.}
\label{fig:misalignment}
\end{figure*}

Ideally, $\phi_{\rm err}=0\rightarrow a=1,\ b=0,\ c=0$, recovering Eq.~\eqref{app:eq:simdriv}. In this case, the angle between the driving direction and the rotating magnetic field is no longer constant over the magnetic-field rotation period \(1/\nu\). As a result, the ratio \(\Omega/\Delta\) becomes time-dependent, and the decoupling condition \(\Omega/\Delta=\sqrt{2}\) is no longer satisfied at all times. Fig.~\ref{fig:misalignment} shows the results for increasing misalignment errors ($\phi_{\rm err}=1^{\circ}$, $\phi_{\rm err}=3^{\circ}$ and $\phi_{\rm err}=5^{\circ}$) obtained with Eq.~\eqref{eq:app:errdriving}. Note that for small angle errors the spectrum is already quite broadened, while for larger misalignments it is completely obscured. 
\section{NV-sample rotation}\label{app:variation}
In this Appendix, we demonstrate that the same results can be obtained is spatially rotating the NV+sample block in stead of the external magnetic field. We then present the numerical simulations under this scenario. For simplicity, let us assume that $\vec{B}_0=\hat{z}$. In this case, the Hamiltonian is 
\begin{align}
H=B_0\sum_k^N\left[\gamma^k\:\vec{I}^k\cdot\left(\mathds{1}+\boldsymbol{\sigma^k}\right)\cdot \hat{z}\right]+\sum_{k<l}^NH_{dd}^{k,l}+H_{\rm c}(t).\label{eq:appF:H}
\end{align}
Since the magnetic field is now static, the RF driving here can be chosen simply as
\begin{align}
        H_c(t)=2\Omega\cos{\left(\omega t+\alpha\right)}\sum_{k=1}^NI_{\hat{x}}^k. \label{appF:eq:simpdriv}
\end{align}
We now follow the same procedure as in Appendix~\ref{app:dyn_ensemble}. In the I.P. with respect to $H_0=\displaystyle\sum_k^N\left[\left(B_0\gamma^k-\Delta^k\right)\ I_{z}^k\right]$, the interaction Hamiltonian is
\begin{align}
    H_I(t)&=\sum_k^N\left[\cos^2{\left(\theta_x^k\right)}\delta_{X_k}+\cos^2{\left(\theta_y^k\right)}\delta_{Y_k}+\cos^2{\left(\theta_z^zk\right)}\delta_{Z_k}\right]I_{z}^k+\sum_{k<l}^N\tilde{H}_{dd}^{k,l}\nonumber\\[6pt]
    &+\sum_k^N\left[\Delta^k\ I_{z}^k-\Omega^k\;I_{\alpha}^k\right],\nonumber
\end{align}
where $I_\alpha=\sin{\alpha}\;I_x+\cos{\alpha}\;I_y$. In this case, $\theta_x^k$,  $\theta_y^k$ and  $\theta_z^k$ come from the explicit form of $R$ in Eq.~\eqref{eq:app:cs_pas}, and represent the angles between each principal axis and the magnetic field. By proceeding as in Appendix~\ref{app:dyn_ensemble}, and choosing now $\alpha=0$, one obtains the same effective dynamics.

\begin{figure*}[h!]
\includegraphics[width=0.5 \linewidth]{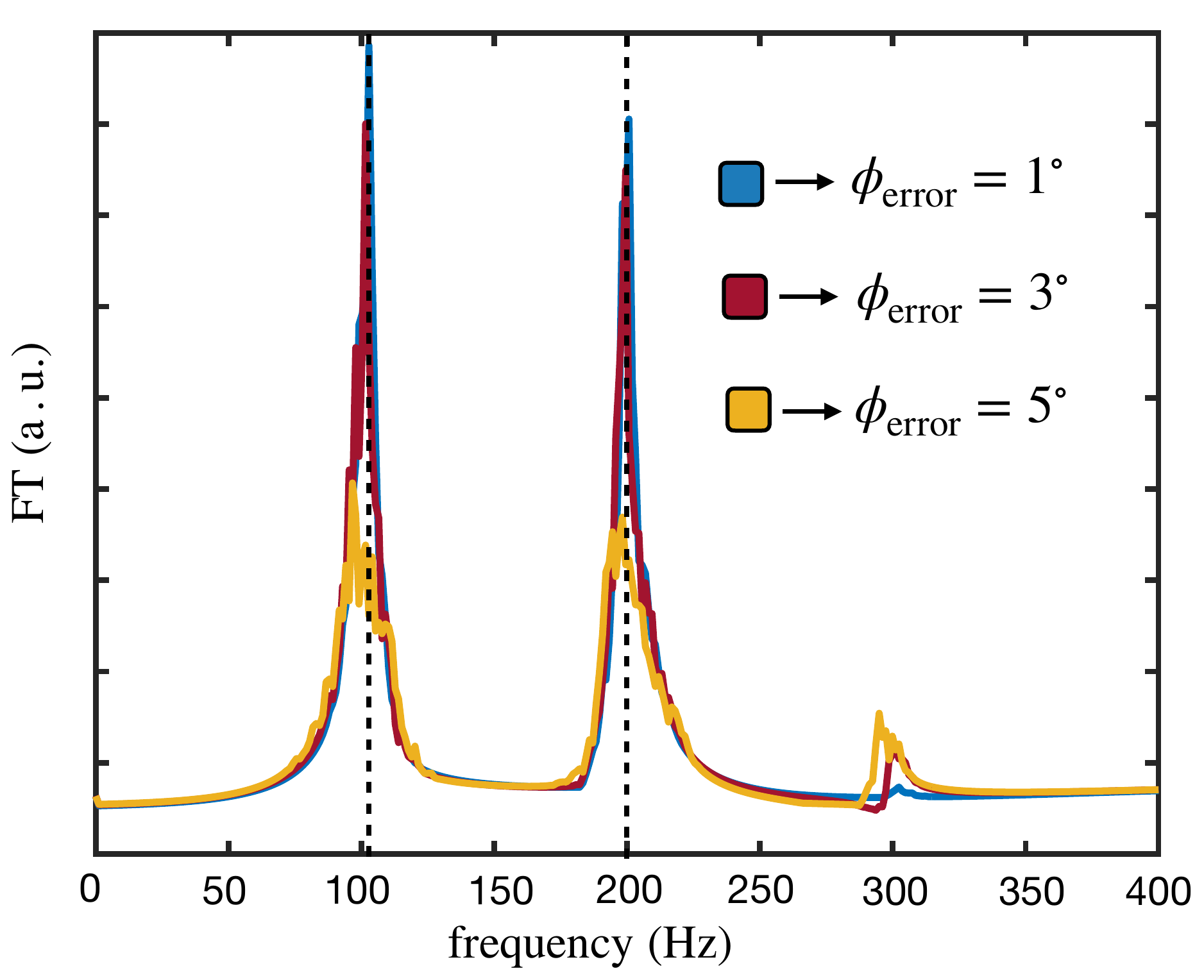}
\caption{Simulated spectrum of the sample with two magnetically nonequivalent protons. In this case, the external magnetic field is fixed, while the sample-NV block is rotated at a rate of $\nu=1$ kHz. The colored solid lines correspond to the simulated spectrum obtained with the version of the driving given by Eq.~\eqref{eq:app:errdriving} for different misalignment errors. Dashed vertical lines indicate the peak position predicted by the theoretical expressions. All the simulations also include stochastic amplitude errors of 0.25\% with correlation time of 1 ms.}
\label{fig:misalignment_spinningsample}
\end{figure*}

The numerical simulations shown in Fig.~\ref{fig:misalignment_spinningsample} show how this protocol leads to the same results and is also robust against misalignment errors.
\end{document}